\documentstyle[twocolumn,aps]{revtex}
\begin{document}

%
%

\title{ Shadow Bands in the Cuprates}

\author{ Stephan Haas, Adriana Moreo, and Elbio Dagotto}

\address{Department of Physics, National High Magnetic Field Lab,
and Supercomputer Computations Research Institute,
Florida State University, Tallahassee, FL 32306, USA}

\date{\today}
\maketitle

\begin{abstract}
A consequence of strong antiferromagnetic correlations in models of
high-Tc cuprates is the appearance in photoemission (PES)
calculations of considerable more weight above the Fermi
momentum ${\rm {\bf p}_F }$ than expected for non-interacting electrons.
This effect, first discussed by Kampf and Schrieffer
(Phys. Rev. {\bf B 41}, 6399 (1990)) under the name of ``shadow bands'', is
here
analyzed in the two dimensional Hubbard and t-J models using Monte Carlo
and exact diagonalization techniques. It is concluded that for
underdoped compounds weight above ${\rm {\bf p}_F }$ could
be observable in experimental PES data, while
in the overdoped regime it
will be likely hidden in the experimental background. In
the intermediate region the signal is weak. Our
results are thus compatible with recent
experimental data by Aebi et al. (Phys. Rev. Lett. {\bf 72}, 2757
(1994)) for Bi2212 at optimal doping.
However, to definitely prove the
existence of this effect in the cuprates, experiments in the
underdoped regime are necessary.

\end{abstract}


%
%

The importance of antiferromagnetic correlations in the normal and
superconducting states of the high critical temperature cuprate
materials is under much discussion. While recently most of the
debate has been concentrated on the symmetry of the superconducting
order parameter, studies of the strength of
the antiferromagnetic correlation length, $\xi_{AF}$, in the normal
state are still crucial to test these ideas. A key issue is how large
$\xi_{AF}$ should be to produce observable effects in experiments for
the high-Tc compounds.
NMR studies in the normal state of optimally doped
${\rm Y Ba_2 Cu_3 O_{6.94}}$ (YBCO)
suggest $\xi_{AF} \sim 2a$ (where $a$ is the lattice spacing).\cite{nmr}
Naively,
this correlation seems too small to be of relevance.
On the other hand, recent photoemission (PES) experimental results by
Aebi et al.\cite{aebi}
in ${\rm Bi_2 Sr_2 Ca Cu_2 O_8}$ (Bi2212) with
Tc=85K, using sequential
angle-scanning data acquisition to obtain PES intensities within a
narrow energy window near the Fermi energy ${\rm E_F}$,
reported evidence compatible with the
antiferromagnetically induced ``shadow bands'' first predicted by Kampf
and Schrieffer.\cite{spinbag}
At half-filling, these bands
are caused by the enlarged magnetic unit cell of the
${\rm CuO_2}$ planes produced by the long range antiferromagnetic
order in the ground state. This effective reduction in the size of the
Brillouin
zone (BZ) has interesting implications for PES experiments.\cite{spinbag} For
example, along the
diagonal ${\rm p_x = p_y = p}$, and assuming long-range order,
peaks at momenta ${\rm {\bf p_1}= (p,p)}$
and ${\rm {\bf p_2} = (\pi - p, \pi - p)}$ should appear at the same energy
location,
for any value of ${\rm p}$. The PES weight observed in the region
$above$ the non-interacting ${\rm {\bf p}_F }$
is induced by strong magnetic correlations.\cite{spinbag}


How does this antiferromagnetically induced PES signal evolve as $\xi_{AF}$ is
made finite by the
effect of hole doping? It is likely
that its intensity will smoothly diminish doping
away from half-filling. Since in optimally doped Bi2212 the antiferromagnetic
correlation length in the ${\rm Cu O_2}$ planes should be similar to that of
optimally doped YBCO (since they should have the same in-plane hole density),
then a natural
question arises: can a small correlation length ($\sim 2a$) produce
observable weight in PES experiments above ${\rm {\bf p}_F}$? On one hand,
recent
calculations\cite{nazarenko} carried out in an antiferromagnetic background
have
reproduced the flat band features near ${\rm {\bf p} = (\pi,0)}$
observed in Bi2212 suggesting that a short $\xi_{AF}$
can appreciably influence some experimental quantities.
On the other hand, since the actual experimental PES signal
(Fig.1a) is weak, concerns may arise
about the interpretation of the data
(as a reference, PES weight induced by antiferromagnetism
in recent experiments for the $insulating$ layered
copper oxide ${\rm Sr_2 Cu O_2 Cl_2}$ is shown in Fig.1b.\cite{wells} ).
Thus, a theoretical calculation is needed to compare with
experiments, and decide if Aebi et al.'s PES data are compatible with models
of strongly correlated electrons which present PES weight above ${\rm
{\bf p}_F}$.

Here, this issue is explicitly addressed.
PES spectra and spin correlations are calculated for models of
correlated electrons expected to qualitatively describe the ${\rm
CuO_2}$ planes. Consider first
the two dimensional (2D) one band Hubbard model defined
by the Hamiltonian
\begin{equation}
{\rm H =-t\sum_{<{\bf i,j}>\sigma} (c^\dagger_{{\bf i}\sigma} c_{{\bf
j}\sigma}+h.c.)
  + U \sum_{\bf i} n_{{\bf i} \uparrow} n_{{\bf i} \downarrow}  },
\end{equation}
\noindent in the
standard notation. We simulated this model numerically using
standard quantum Monte Carlo
techniques. In order to extract the
dynamical spectral function ${\rm A({\bf p},\omega)}$ corresponding
to the removal or
addition of an electron with momentum ${\bf p}$ to the system, the
maximum entropy (ME)
technique was used.\cite{ME} To analyze the strength of the
signal above ${\rm {\bf p}_F}$,
we calculated the amount of spectral weight below the chemical potential
$\mu$ at momenta along the diagonal ${\rm p_x = p_y}$ in the BZ with
respect to the total
intensity (adding
PES and inverse PES) which for the Hubbard model satisfies the sum rule
$\int^{+\infty}_{-\infty} d\omega A({\bf p},\omega) = 1$
at all dopings.

Before describing the computational results, let us
clarify when a theoretically calculated PES signal can be labeled as being
``observable'' in an experiment. PES spectra have large backgrounds,
whose origin and shape are a matter of discussion,
superimposed on the actual relevant signal. This background depends on the
momentum,
and also changes from sample to sample with fluctuations as
large as 50\%. Since the background is convex,
the natural requirement
for a PES theoretical signal to be observable is that the combination
background-signal produces a local maximum (i.e. a peak in the measured
intensity).\cite{wells2}  From the data shown in Fig.1b, and the intensity of
the
signal at the last point where the dispersion is observed i.e. ${\bf p} = (0.7
\pi, 0.7 \pi)$, we believe that a peak with an intensity of
roughly about 10\%
of the largest signal (located at ${\bf p} = (0.5 \pi, 0.5 \pi)$)
would be at the verge of being detected.\cite{wells2} This is the
criterion followed in the present paper to label a result as ``observable''.

The choice of coupling is important in our search
for PES weight above ${\rm {\bf p}_F}$.
For example, we observed that working on an $8\times8$ cluster, at
${\rm U/t=4}$, half-filling and temperature ${\rm T=t/4}$, the
percentage of PES spectral weight at ${\rm {\bf p} = (3\pi/4, 3\pi/4)}$,
i.e. the next available momentum after ${\rm (\pi/2,\pi/2)}$ on this cluster,
is very small (less than 5\% of the total), even though the spin correlations
show
clear indications of long-range order. Then, the actual value of the
local moments is as important as the antiferromagnetic correlation length
for the effect we are investigating.
Since evidence has been recently given
that another feature induced by antiferromagnetism,\cite{duffy} namely the
``hole''
pockets, are washed out by temperature effects in QMC simulations at ${\rm
U/t=4}$,
then we consider this coupling to be too small for our purposes.
Actually,
studies of the optical conductivity have shown that a larger coupling
${\rm U/t}$, approximately between 8 and 12, is needed to correctly reproduce
the main
features observed experimentally.\cite{review} Since
at ${\rm U/t=12}$ there are serious numerical
instabilities in the simulations,
then here the analysis was restricted to ${\rm U/t=8}$.


In Fig.2, ${\rm A({\bf p},\omega)}$ at ${\rm T=t/2}$ is shown.\cite{bulut} It
is
difficult to reduce ${\rm T}$ due to sign problems, but
nevertheless this temperature allows us to study the PES signal
above ${\rm {\bf p}_F}$ at
different correlation lengths as the density is changed, which is the
main purpose of the paper.
At half-filling, ${\rm \langle n \rangle = 1}$, the chemical potential is
located in the
gap. The percentage of spectral weight is shown for each momentum.
A nonzero signal above the non-interacting Fermi momentum
is clearly visible, and at ${\rm
{\bf p}=(3\pi/4,3\pi/4)}$
it carries ${\rm \sim 23\%}$ of the total weight. This
result is very similar if the temperature is reduced to ${\rm T=t/4}$,
thus finite temperature effects are
not too severe for this quantity at half-filling. Actually, our results for
the intensity of the weight above ${\rm {\bf p}_F}$ are
in excellent agreement with the spin density wave mean-field
approximation.\cite{ortolani}
Since for a pure spin-1/2 antiferromagnet the weight at
${\rm {\bf p}=(\pi/4,\pi/4)}$
and ${\rm {\bf p}=(3\pi/4,3\pi/4)}$ should be identical at this density,
it is natural to conclude that the finite coupling ${\rm U/t}$ is responsible
for the
reduction of the intensity of the PES signal above the naive ${\rm {\bf
p}_F}$ at half-filling.

Away from half-filling, at $\langle n \rangle = 0.87$, the amount of
weight at ${\rm {\bf p}=(3\pi/4,3\pi/4)}$ is reduced to ${\rm \sim
10\%}$, which is still visible in the scale of the plot.
The height of the peak, as a percentage of the peak height at ${\rm {\bf
p} = (\pi/2,\pi/2) }$ and half-filling
is about 15\%.
Following the criteria described before, the result
obtained at ${\rm \langle n \rangle = 0.87}$ in the Hubbard model is labeled
as still ``observable'' (although it is rather
weak). Whether this PES weight corresponds to an actual sharp band dispersing,
to a broader feature, or a combination of both,
is difficult to address with the ME
technique which has low resolution,
but nevertheless it is clear that it is induced by
antiferromagnetism.
At $\langle n \rangle = 0.70$, the signal at ${\rm {\bf
p}=(3\pi/4,3\pi/4)}$  carries a small weight
of only 4\%, and as the electronic density is reduced further  the system
smoothly converges to the non-interacting limit.
Then, for this particular calculation we tentatively
conclude that a doping of 25\% holes makes the
antiferromagnetically induced weight
almost negligible, while at 12\% doping the effect is still
observable.
To relate these results with experiments, in Fig.3a we show the
spin correlations calculated numerically.
At $\langle n \rangle =1$,
the correlation is robust but it
decays slowly to zero due to temperature effects, while
at $\langle n \rangle =0.70$ it is clearly very small. At an intermediate
density
$\langle n \rangle = 0.87$, $\xi_{AF}$ is between one and two lattice
spacings, resembling the experimental situation in YBCO, and presumably
also in Bi2212, since both are at optimal doping.\cite{comm10}
In spite of this fact,
the weight above ${\rm {\bf p}_F}$
is observable in Fig.2, showing that the effect is mainly
caused by
the local short-distance antiferromagnetic spin arrangement. Thus, the
presence of large spin correlation lengths
is $not$ a necessary condition for the observation of
weight above the noninteracting Fermi momentum.


To check the sensitivity of our conclusions to the model used, let us also
consider the
well-known 2D ${\rm t-J}$ Hamiltonian.
This model cannot be studied with
Monte Carlo techniques, thus exact diagonalization was
used.\cite{review} This algorithm works at T=0,
and dynamical information can be obtained in real time. Its
restriction to small clusters should not be a major problem
in calculations where $\xi_{AF}$ is very small.
To increase the momentum resolution
along the diagonal in the BZ, we combined the results of the
16 sites cluster (providing momenta ${\rm (0,0),
(\pi/2,\pi/2),(\pi,\pi)}$)
and the 18 sites cluster (containing ${\rm (\pi/3,\pi/3),
(2\pi/3,2\pi/3)}$).\cite{foot2}

In Fig.4, the PES ${\rm A({\bf p},\omega)}$ spectrum is shown for the t-J
model.
${\rm J/t=0.4}$ was selected as
an example, but we checked that the results are qualitatively similar in
the range
between ${\rm J/t=0.2}$ and ${\rm J/t=0.8}$. As expected,
at half-filling the
largest peak near the chemical potential (quasiparticle) is obtained
at ${\rm {\bf p} =(\pi/2,\pi/2)}$. Increasing the
diagonal momenta away from it, a considerable amount of spectral weight
induced by $\xi_{AF}$ is observed. Moving
away from half-filling into the subspace of two holes (nominal
density $\langle n \rangle \sim 0.88$) the dominant peak remains
at ${\rm {\bf p}=(\pi/2,\pi/2)}$ within our momentum
resolution. At $(\pi/3, \pi/3)$
the quasiparticle
strength is still large and coherent. On the other hand,
at ${\rm {\bf p}=(2\pi/3,2\pi/3)}$ the peak seems now broader in the
scale used, although
its integrated spectral weight remains close to that of
${\rm {\bf p}=(\pi/3,\pi/3)}$. The height of the peak at
${\rm {\bf p}=(2\pi/3,2\pi/3)}$ as a percentage of the largest peak
located
at ${\rm {\bf p}=(\pi/2,\pi/2)}$ (with or without holes)
is 15-20\% i.e. within the ``observable''
region defined before.
Finally, at density $\langle n \rangle =
0.77$, the result resembles that of a non-interacting system
with a Fermi momentum close to
${\rm {\bf p}=(\pi/3,\pi/3)}$, above which the signal is too weak to be
observable in PES experiments.
Then, our rough estimations within the t-J model are similar to those of
the Hubbard model, i.e. weight above ${\rm {\bf p}_F}$
can still be observed at $\langle n \rangle \sim 0.88$ but no longer at density
$\langle n \rangle \sim 0.77$. To make contact with experiments it is again
necessary to consider the spin correlations
which are shown in Fig.3b.
At half-filling,
$\xi_{AF}$ is clearly larger than the lattice size. At $\langle n
\rangle \sim 0.88$, a crude exponential fit of the
spin correlation vs. distance
gives $\xi_{AF} \sim 1.5 a$ (similar to that of
YBCO and Bi2212 at optimal doping), while at $\langle n
\rangle \sim 0.77$, $\xi_{AF}$ is less than one lattice spacing.
Then, we again arrive to the conclusion that for a real material with
$\xi_{AF} \sim 2a$
the antiferromagnetically generated PES weight is
weak but may still be observable above the background.

In Fig.5, ${\rm A({\bf p},\omega) }$ is shown again
at $\langle n \rangle \sim 0.88$ using an expanded energy scale.
The dispersion of the sharp peak (I) discussed before
in Fig.4, has a bandwidth of order J, while at higher energies
a considerable amount of spectral weight is found contributing to
the bulk of the valence band (II). This double peak structure is in good
agreement with the prediction of Kampf and Schrieffer\cite{spinbag} i.e. the
lowest
energy peak (I) and the incoherent part (II)
at ${\rm {\bf p} = (2 \pi/3, 2\pi/3) }$ are the shadow bands caused by
antiferromagnetic correlations.
ME entropy techniques cannot resolve these two peak
structure, and thus exact diagonalization is needed to quantitatively analyze
these features.

Summarizing, an analysis of the PES spectra in the
2D Hubbard and t-J models, at several
densities and couplings, was reported. If these models qualitatively reproduce
the physics
of the high-Tc compounds, then we conclude that the
effect predicted by Kampf and Schrieffer based on a robust $\xi_{AF}$
should be observable even for
materials with spin correlations lengths of only a couple of
lattice spacings, as in Bi2212 at optimal doping. This is $compatible$ with the
experimental
results of Fig.1a. However, this regime
is at the verge of observability.
The PES signal above ${\rm {\bf p}_F}$
should no longer be visible above the large PES experimental
background at slightly larger dopings.
To gather further evidence that
the weak experimental signal is indeed caused by antiferromagnetism
we believe that it is necessary to carry out PES experiments as a function
of hole doping. The strength of the signal above ${\rm {\bf p}_F}$
should increase as the
system moves away from the optimal doping towards
half-filling. A possible candidate for such a
study is YBCO with a critical temperature of about 60K.
Another alternative within the Bi2212 family would be to consider
${\rm Bi_2 Sr_2     Ca_{1-x} Lu_x Cu_2 O_{8+\delta}}$ and
${\rm Bi_2 Sr_{2-x} La_x     Ca   Cu_2 O_{8+\delta}}$ which seem to be
underdoped.\cite{under}

\medskip
We thank J. R. Schrieffer, P. Monthoux, D. Scalapino, M. Onellion,
and A. Nazarenko for conversations,
and J. Riera for providing us with programs for
the t-J model.
E. D. and A. M. are supported by the Office of Naval Research under
grant ONR N00014-93-0495. E. D. also thanks the donors of the
Petroleum Research Fund administered by the American Chemical Society.
We also thank MARTECH (FSU) for support.
\medskip

\vfil\eject

%
%

{\bf Figure Captions}

\begin{enumerate}

\item (a) PES intensity
in Bi2212, as reported by Osterwalder et al. (Ref.\cite{aebi}). Their
method produces PES intensity at constant energy for all momenta,
while conventional methods provide
complete PES energy distribution curves at a few locations in
the BZ. Each solid line corresponds to a fixed energy scan
starting at the bottom at 0.3 eV above ${\rm E_F}$, and arriving at the last
top line at ${\rm E_F}$. The spectra were
measured at a polar angle of ${\rm 39^0 }$,
and for azimuthal angles spaced
${\rm 1^0}$ apart beginning near the ${\rm \Gamma M}$ line and ending near the
${\rm \Gamma X}$ line. The ``$5\times1$'' band is explained in the
original text Ref.\cite{aebi} and it is of no concern to us.
We thank P. Aebi and J. Osterwalder for providing us with these
unpublished data; (b) PES intensity as a function of energy for
${\rm Sr_2 Cu O_2 Cl_2}$ taken from B. O. Wells et al.
(Ref.\cite{wells}).
The momenta are given in units of $\pi$ and along the
diagonal in the 2D square lattice convention.


\item
${\rm A({\bf p},\omega) }$, evaluated using QMC and ME
techniques, for the 2D Hubbard model at
${\rm U/t=8}$, ${\rm T = t/2}$ on an $8\times 8$ cluster, at several
densities ${\rm \langle n \rangle }$. The momentum label varies along the
diagonal in the BZ in units of $\pi/4$, and the percentages
correspond to the integrated PES
part of the spectral weight with respect to the total intensity (=1).
The energy is in units of t.

\item
(a) Spin-spin correlation ${\rm 4 \langle S^z_{\bf i} S^z_{\bf i+j}
\rangle (-1)^{|{\bf j}|} }$ vs distance, ${\rm j = |{\bf j}| }$, for the 2D
Hubbard model
calculated using QMC at ${\rm T=t/2}$,
${\rm U/t=8}$, and several densities on an $8 \times 8$ cluster;
(b) Spin-spin correlations as defined in Fig.3a, for the 2D t-J model
calculated using exact diagonalization techniques on a $4\times 4$
cluster with two holes, at several couplings.

\item
PES ${\rm A({\bf p},\omega) }$ evaluated using
exact diagonalization techniques for
the 2D t-J model, at ${\rm J/t=0.4}$ on $4 \times 4$ and
$\sqrt{18}\times \sqrt{18}$ clusters. The densities are shown in the
figure. We assumed ${\rm t= 0.4 eV}$, and provided a width ${\rm \delta =
0.1t}$ to the peaks.
The momenta are indicated, and the relevant peaks are
shaded.

\item
PES ${\rm A({\bf p},\omega) }$ for the t-J model at
$\langle n \rangle \sim 0.88$, ${\rm J/t=0.4}$, clusters of 16 and 18
sites, and expanding the energy
scale to observe the two peak structure. We use $\delta = 0.25t$ and
${\rm t=0.4 eV}$.

\end{enumerate}


\begin{references}
\bibitem{comm10} Here, we $assume$ that the PES spectral weight induced by
antiferromagnetism is mainly a function of $\xi_{AF}$ i.e. we compare
numerical and experimental results at different temperatures,
but having the same spin correlation.
\bibitem{foot2} These calculations were carried out on a
Cray-90 supercomputer and IBM workstations.
The largest Hilbert
space handled (i.e. N=18, 5 holes)
contains 816816 states, exploiting the translational invariance symmetry.
Typically each PES figure contains about 200 poles.
\end{references}
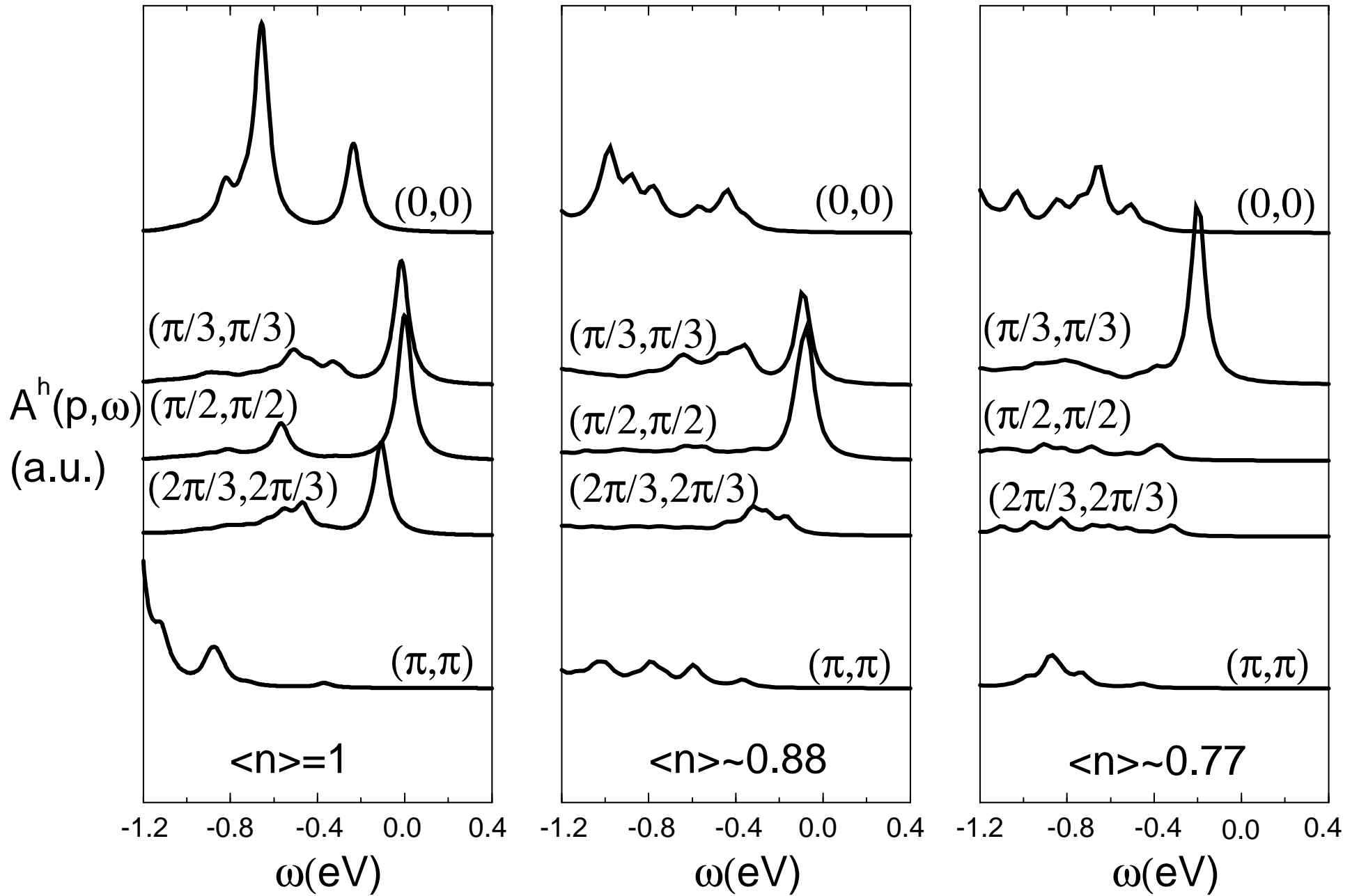
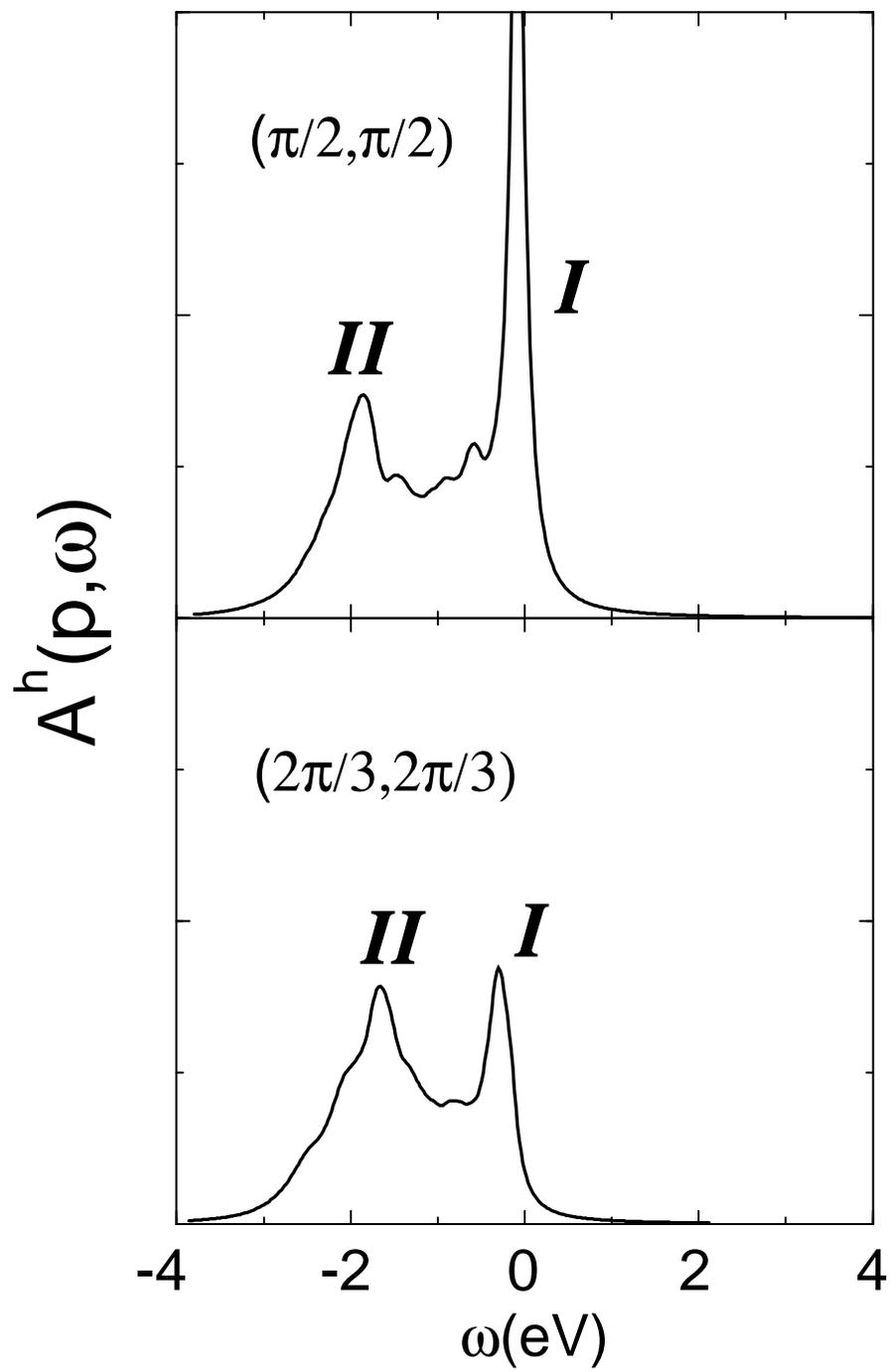
\end{document}